\documentclass[twocolumn]{aastex631}

\usepackage{makecell}
\usepackage{enumerate}
\usepackage{graphicx}
\usepackage{float}
\usepackage{subfigure}
\usepackage{amsmath}
\usepackage{CJKulem}
\usepackage{threeparttable}
\usepackage{footnote}
\usepackage{appendix}
\begin{document}
\shortauthors{Hu et al.}

\title{Beam pattern evolution of accreting X-ray pulsar 1A 0535+262 during its 2020 giant outburst}

\correspondingauthor{L. Ji}
\email{jilong@mail.sysu.edu.cn}

\author{Y.F. Hu}
\affiliation{School of Physics and Astronomy, Sun Yat-Sen University, Zhuhai, 519082, People’s Republic of China}
\author{L. Ji}
\affiliation{School of Physics and Astronomy, Sun Yat-Sen University, Zhuhai, 519082, People’s Republic of China}
\author{C. Yu}
\affiliation{School of Physics and Astronomy, Sun Yat-Sen University, Zhuhai, 519082, People’s Republic of China}
\author{P.J. Wang}
\affiliation{University of Chinese Academy of Sciences, Chinese Academy of Sciences, Beijing 100049, People’s Republic of China}
\affiliation{Key Laboratory of Particle Astrophysics, Institute of High Energy Physics, Chinese Academy of Sciences, Beijing 100049, People’s Republic of China}
\author{V. Doroshenko}
\affiliation{Institut für Astronomie und Astrophysik, Kepler Center for Astro and Particle Physics, Eberhard Karls, Universität, Sand 1, D-72076 Tübingen, Germany}
\author{A. Santangelo}
\affiliation{Institut für Astronomie und Astrophysik, Kepler Center for Astro and Particle Physics, Eberhard Karls, Universität, Sand 1, D-72076 Tübingen, Germany}
\author{I. Saathoff}
\affiliation{Institut für Astronomie und Astrophysik, Kepler Center for Astro and Particle Physics, Eberhard Karls, Universität, Sand 1, D-72076 Tübingen, Germany}
\author{S.N. Zhang}
\affiliation{University of Chinese Academy of Sciences, Chinese Academy of Sciences, Beijing 100049, People’s Republic of China}
\affiliation{Key Laboratory of Particle Astrophysics, Institute of High Energy Physics, Chinese Academy of Sciences, Beijing 100049, People’s Republic of China}
\author{S. Zhang}
\affiliation{Key Laboratory of Particle Astrophysics, Institute of High Energy Physics, Chinese Academy of Sciences, Beijing 100049, People’s Republic of China}
\author{L.D. Kong}
\affiliation{University of Chinese Academy of Sciences, Chinese Academy of Sciences, Beijing 100049, People’s Republic of China}
\affiliation{Key Laboratory of Particle Astrophysics, Institute of High Energy Physics, Chinese Academy of Sciences, Beijing 100049, People’s Republic of China}

\begin{abstract}
We report on pulse profile decomposition analysis of a bright transient X-ray pulsar 1A~0535+262 using the broadband {\it Insight}-HXMT observations during a giant outburst of the source in 2020. We show that the observed  pulse profile shape can be described in terms of a combination of two symmetric single-pole contributions for wide range of energies and luminosities for a fixed geometry defining basic geometry of the pulsar. This corresponds to a slightly distorted dipole magnetic field, i.e., one pole has to be offset by $\sim 12^{\circ}$ from the antipodal position of the other pole. We reconstruct the intrinsic beam patterns of the pulsar assuming the geometry recovered from the decomposition analysis, and find evidence for a transition between ``pencil" and ``fan" beams in energy ranges above the cyclotron line energy which can be interpreted as transition from sub- to super-critical accretion regimes associated with onset of an accretion column. At lower energies the beam pattern appears, however, to be more complex, and contains substantial ``fan" beam and an additional ``pencil" beam component at all luminosities. The latter is not related to the accretion rate and is stronger in the fading phase of the outburst. We finally discuss results in context of other observational and theoretical findings earlier reported for the source in the literature.

\end{abstract}

\section{Introduction} \label{Introduction}
In high mass X-ray binaries (HXMBs), compact objects accrete material from a companion star with a mass greater than 10 solar masses via the Roche lobe or winds \citep{Iben1991ApJS...76...55I, Davidson1973ApJ...179..585D,Frank1992apa..book.....F}.
If the compact star is a highly magnetized neutron star, the accreted matter will be channelled by its magnetic field onto the magnetic poles on the surface of the compact object ultimately converting gravitational potential energy into X-rays, which can be pulsed if the spin axis and the magnetic axis of the neutron star are misaligned. 

Pulse profiles of accreting X-ray pulsars are known to have complex morphology, which is in general highly variable with energy and the luminosity \citep[e.g.,][]{AlonsoHernandez2022}.
Observed changes of pulse profile shapes with energy is believed to reflect details and angular dependence of radiation transfer in the emission region and often referred to as the intrinsic beam pattern of a pulsar\footnote{In this paper, the "beam pattern" refers to the flux distribution as a function of the angle between the dipole magnetic axis and the line of sight to a distant observer.
The "intrinsic beam pattern" is the beam pattern without considering the relativistic light deflection.}, while variations with luminosity indicate the changes of the emission region geometry with the accretion rate \citep[see, e.g.,][]{Mushtukov2022}.

In order to interpret the observed pulse profiles, several theoretical models were proposed \citep[][]{Wang1981A&A...102...97W, Meszaros1985, Ferrigno2011, Cappallo2017}.
However, details of interaction of magnetic field, light and matter near the polar caps are still poorly understood, which hampers detailed and physically motivated modeling of observed pulse profiles.
An alternative approach to the direct modeling of pulse profiles has been proposed by \cite{Kraus1995ApJ...450..763K} where the intrinsic beam pattern associated with each individual magnetic pole is assumed to be intrinsically symmetric. The asymmetric shape of observed pulse profiles is then attributed to offset of the magnetic dipole from the center of the pulsar. \cite{Kraus1995ApJ...450..763K} exploit energy and luminosity variations of the observed pulse profiles to find a unique solution defining geometry of the pulsar, which needs to remain constant regardless on details of radiative transfer or changes of emission region geometry with accretion rate. This decomposition method has been successfully applied to several accretion pulsars, such as Cen X-3, Her X-1, EXO 2030+375, 1A 0535+262, 4U 0115+63 and V 0332+53 \citep{Kraus1996ApJ...467..794K, Blum2000ApJ...529..968B, Sasaki2010A&A...517A...8S, Caballero2011A&A...526A.131C, Sasaki2012A&A...540A..35S}. The advantage of the method is that interpretation of the reconstructed intrinsic beam patterns associated with individual poles is arguably easier than full modeling of the pulse profiles from theoretical perspective.
It is also worth noting that Her X-1's geometry estimated by the pulse profile decomposition \citep{ Blum2000ApJ...529..968B} appears to be consistent with the recent polarization study performed by the Imaging X-ray Polarimetry Explorer (IXPE) \citep{Doroshenko2022}. This indicates that the decomposition method might indeed be a good approximation for reconstructing the beam pattern of pulsars.

The transient X-ray pulsar 1A~0535+262 was discovered by Ariel V during a giant outburst in 1975  \citep{Rosenberg1975Natur.256..628R}.
It consists of a pulsating neutron star with a spin period of 103\,s and a Be companion HD~245770 \citep{Hudec1975MitVS...7...29H}.
The distance to the source is estimated at about 2\,kpc measured by Gaia \citep{Bailer-Jones2018AJ....156...58B}. 1A 0535+262 is an active Be X-ray binary that has shown frequent outbursts in its history \citep[see][and references therein]{CameroArranz2012}.
\citet{Caballero2011A&A...526A.131C} applied the decomposition method to the pulse profiles observed during the outburst in 2005 with \textit{RXTE} mission at luminosity level of about $\rm 0.8\times 10^{37}\,erg\,s^{-1}$.  They estimated the polar angles defining geometry of the pulsar $\Theta_1 \approx 50^\circ$ and $\Theta_2 \approx 130^\circ$ (see below), and found a possible solution of the beam pattern interpreted as a hollow column plus a halo of radiation scattered off the neutron star surface.

In November 2020, the brightest outburst ever recorded from the source was observed, with a luminosity reaching $\rm 1.2\times 10^{38}\,erg\,s^{-1}$. \citet{Kong2021ApJ...917L..38K} found that the observed energy of the cyclotron absorption line was anti-correlated with the luminosity around the burst peak, which is an important change compared to historical observations at lower fluxes  where no such correlation was observed, and which can be interpreted as an evidence for the transition of accretion regimes. \citet{Kong2021ApJ...917L..38K} also reported significant differences in observed broadband X-ray spectrum between the rising and fading phases of the outburst, suggesting a somewhat different emission region geometry even at the same accretion rate in two phases of the outburst.
On the other hand, using \textit{Insight}-HXMT observations, \citet{Wang2022} reported the complex pulse profile evolution throughout the outburst, which exhibits a strong energy and luminosity dependence. Here we investigate observed variations of the pulse profile more quantitatively and apply decomposition method by \cite{Kraus1995ApJ...450..763K} to \textit{Insight}-HXMT  data. This allows us to recover the evolution of intrinsic beam patterns as a function of luminosity (including both rising and fading phases) in 1A 0535+262 including also the previously unexplored range of luminosities close to the peak of the outburst. The paper is organized as follows: Sections~\ref{Data} and~\ref{method} briefly describe the data and the decomposition method used in this paper. Results are presented in Section~\ref{results}. Finally, we discuss results and present conclusions in Section~\ref{discussions}.

\section{Available data and data selection}\label{Data}
\begin{figure*}[!htp]
\vspace{-0.7cm}
\gridline{\fig{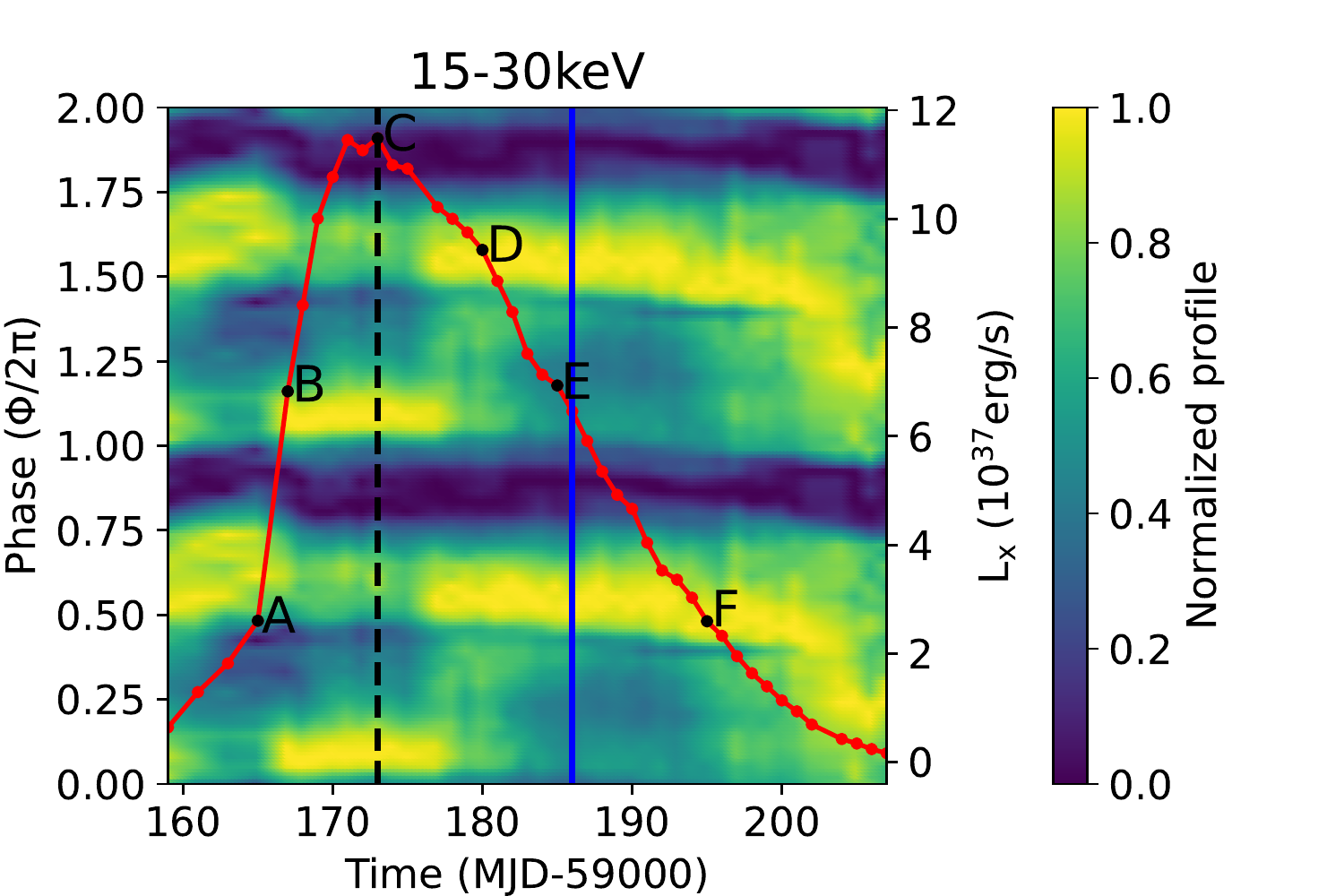}{0.463\textwidth}{}
          \fig{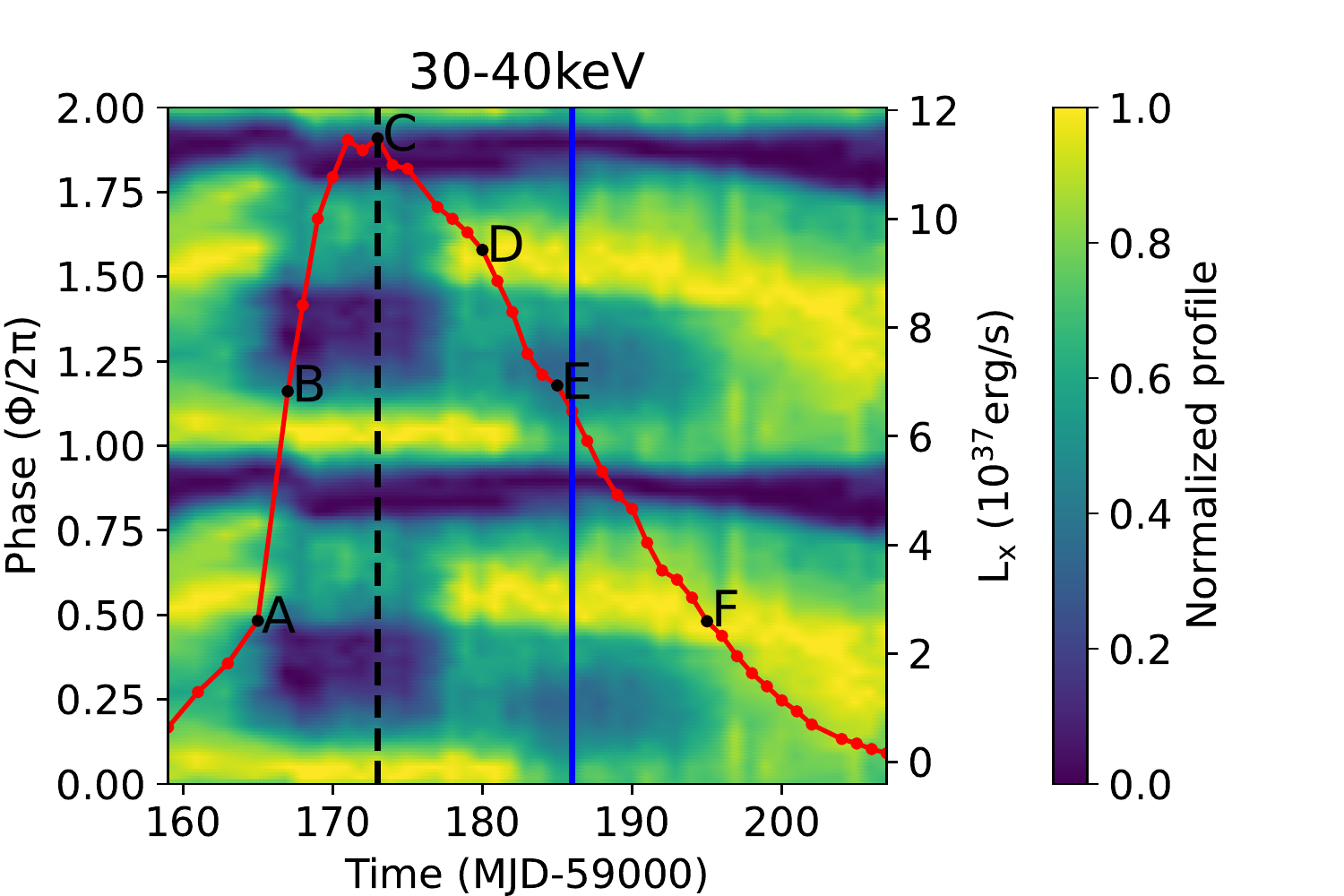}{0.463\textwidth}{}}\vspace{-0.7cm}
\gridline{\fig{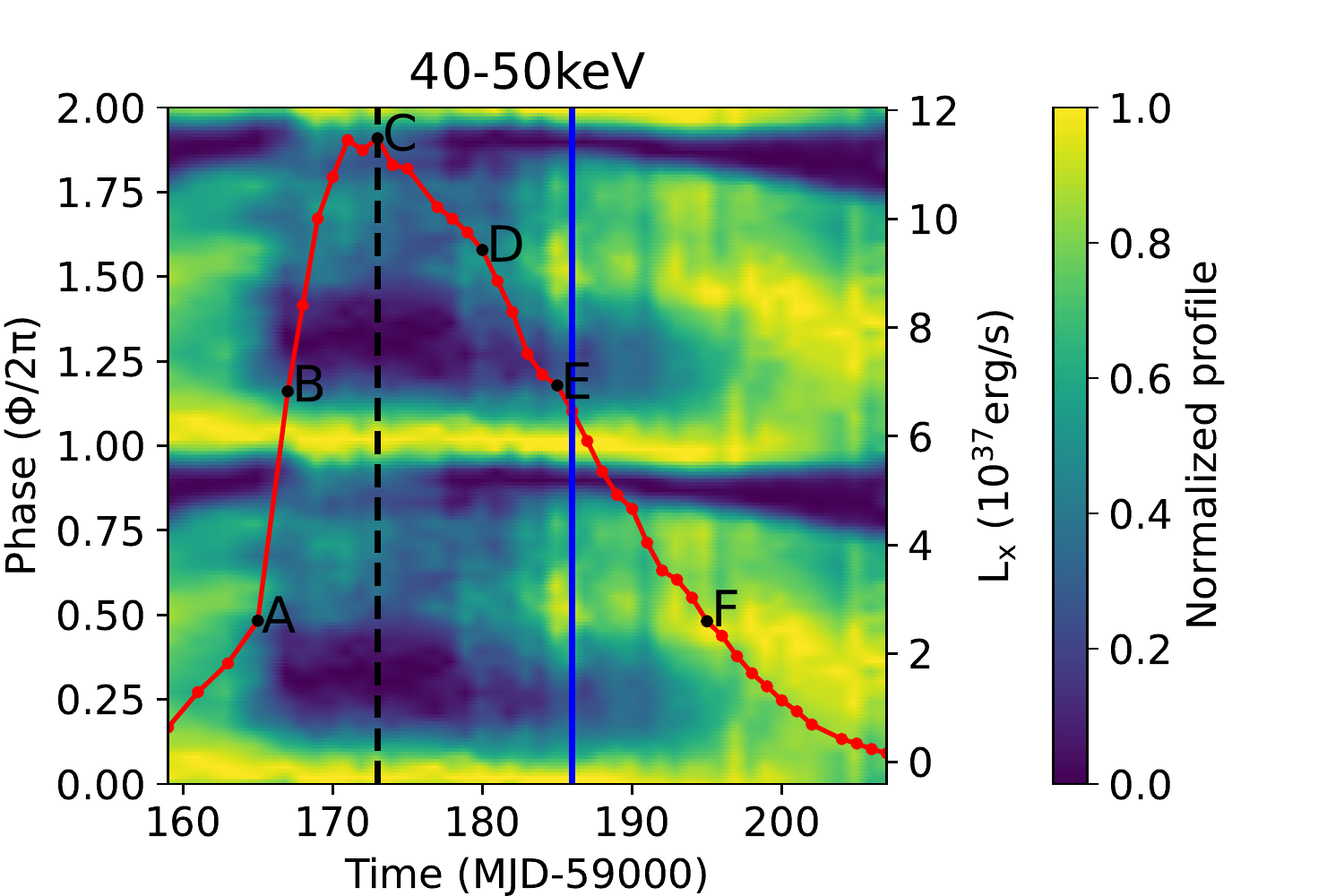}{0.463\textwidth}{}
          \fig{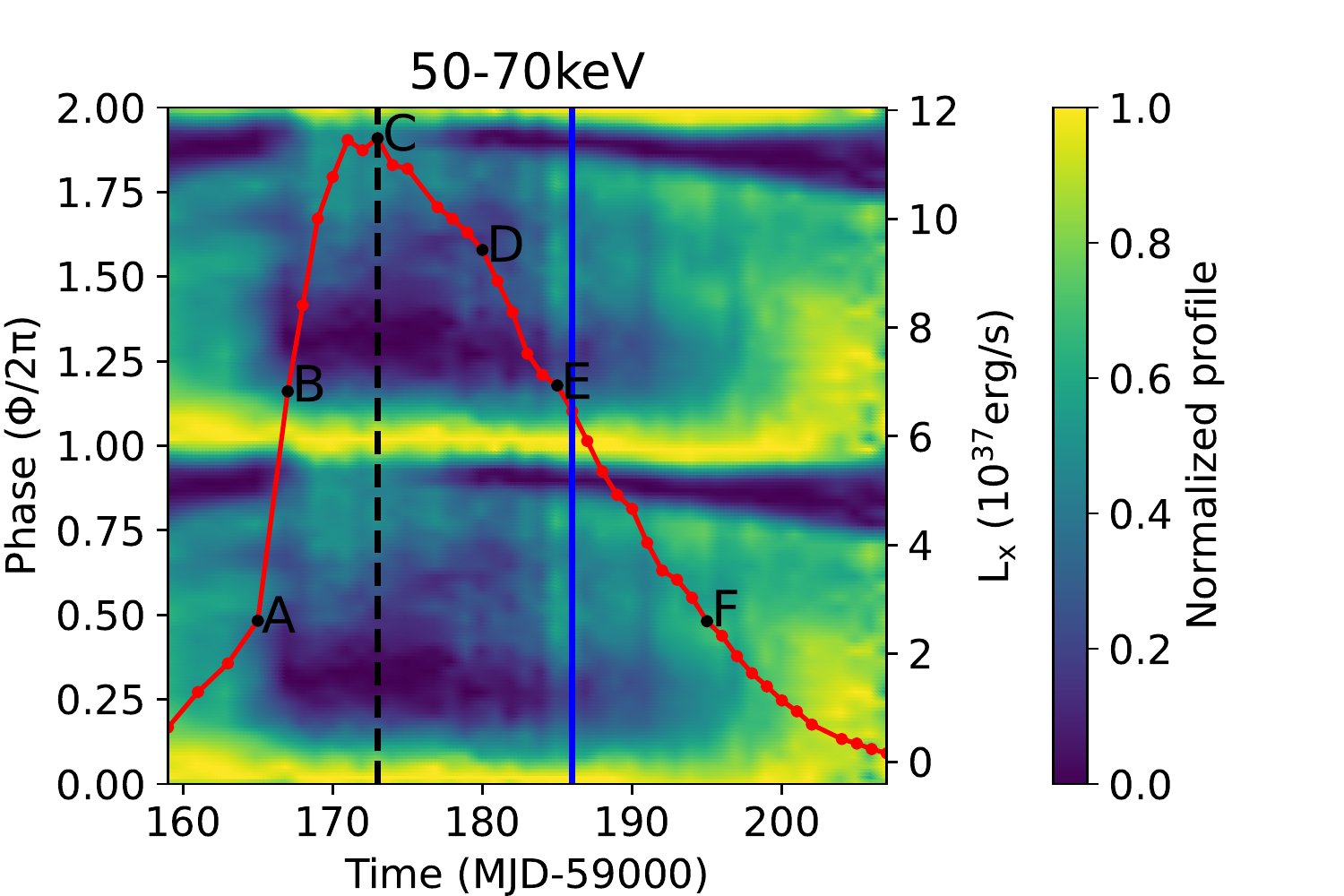}{0.463\textwidth}{}}\vspace{-0.7cm}
%
\caption{The evolution of the pulse profiles in the energy ranges of 15-30\,keV, 30-40\,keV, 40-50\,keV and 50-70\,keV, respectively. The colors present the intensity of pulse profiles normalized in the [0,1] range.
The long-term lightcurve is shown as the red line in each panel. 
Six representative observations (A-F) are selected for the following decomposition.
The black dashed vertical line indicates the outburst peak and the blue line corresponds to the time when the source has the critical luminosity in the decay phase of the outburst.
}
\vspace{-0.0cm}
\label{flux}
\end{figure*}

\textit{Insight}-HXMT \citep{Zhang2014SPIE.9144E..21Z,Zhang2020SCPMA..6349502Z} performed a high-cadence observational campaign and obtained unprecedented high quality data of 1A 0535+026 during its giant outburst in 2020. In Figure~\ref{flux} we show observed evolution of the pulse profiles (adopted  from \citet{Wang2022}), as a function of bolometric luminosity of the source estimated using the broadband spectroscopy in the energy range of 2-150\,keV and assuming a distance of 2\,kpc. During the outburst, as reported by \citet{Wang2022}, the pulse profile shows a complex variation with energy and luminosity. This is an important prerequisite for our application of the decomposition method \citep{Kraus1995ApJ...450..763K}.

The energy ranges we considered in this study is 15-30\,keV, 30-40\,keV, 40-50\,keV and 50-70\,keV. This is due to the fact that a significant fraction of low energy photons come from thermal components \citep{Kong2021ApJ...917L..38K} and they may not be emitted directly from the polar caps \citep{Poutanen2013, Tao2019}. On the other hand, at higher energies the radiation is dominated by the instrumental background. In addition, the centroid energy ($E_{\rm cyc}$) of cyclotron resonant scattering features (CRSFs) in 1A 0535+262 is approximately 45\,keV and the energy-dependent cross section results in dramatic changes of pulse profiles around $E_{\rm cyc}$ (see Figure~4 in \citet{Wang2022}). The energy ranges we used are due to a trade-off between statistics and pulse profile variation with energy. The details of data reduction and analysis are presented in \citet{Wang2022}, and here we focus on the decompositional analysis of the obtained pulse profiles in several energy ranges.

To investigate possible changes of the emission region geometry, we decomposed pulse profiles observed at several characteristic luminosities marked in Figure~\ref{flux} (A-F). Point C is at the outburst peak with a luminosity $L \approx \rm 11.5\times 10^{37}\, ergs\,s^{-1}$. Point E corresponds to the critical luminosity ($L\approx \rm 6.7\times 10^{37}\, ergs\,s^{-1}$) proposed by \citet{Kong2021ApJ...917L..38K}, above and below which the accretion regimes are expected to be different \citep[see, e.g.,][]{Basko1976MNRAS.175..395B, Becker2012, Mushtukov2015}. In addition, as reported by Figure~2 in \cite{Wang2022}, a transition of pulse profiles in 10-30\,keV appears when $L \approx \rm 9.5\times 10^{37}\, ergs\,s^{-1}$, and therefore, point D is included in this study. Point F is also taken into account to represent a relatively low luminosity state ($L \approx \rm 2.6\times 10^{37}\, ergs\,s^{-1}$) \footnote{Another transition of pulse profiles that occurs at $L \approx \rm 1.1\times 10^{37}\, ergs\,s^{-1}$  \citep{Wang2022} is not included in the paper. This is because 1) the analysis at a similar luminosity state has been done by \citet{Caballero2011A&A...526A.131C}; 2) the pulse phase at low luminosities can not be well aligned with those at high luminosities, due to the sudden change of pulse profiles; 3) at low luminosities uncertainties of the background estimation significantly influence the "non-negative" criterion (see text below).
}. Finally, points A and B, which have almost the same luminosities compared with points F and E, are also included to assess possible differences in intrinsic beam patterns between rising and fading phases of the outburst.

\begin{figure*}
  \centering
  \includegraphics[width=0.8\textwidth]{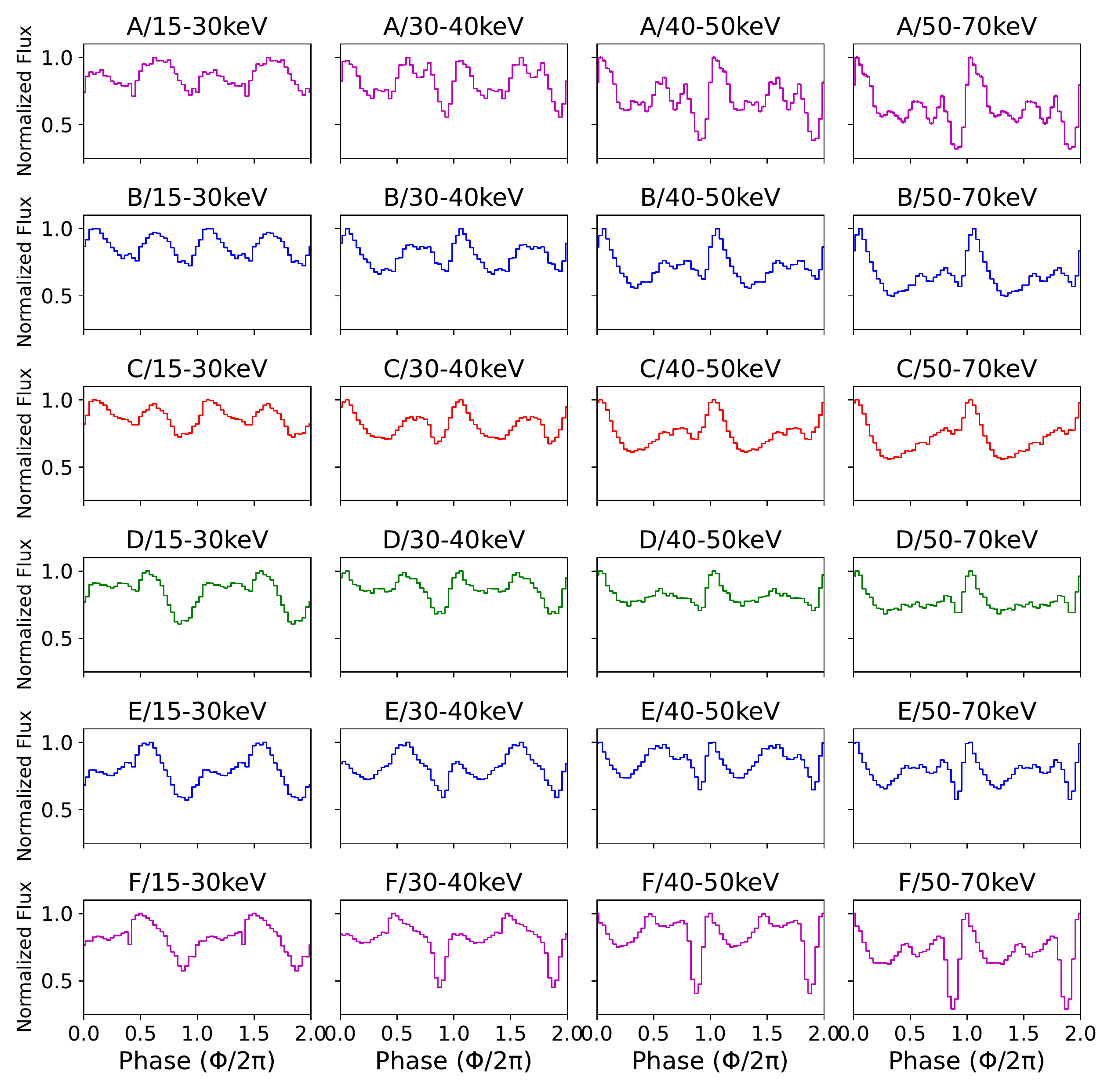}
\caption{Pulse profiles of 1A 0535+262 during its 2020 giant outburst observed by Insight-{\it HXMT}. All pulse profiles were normalized to their maximum values.}
\vspace{-0.0cm}
\label{profile1}
\end{figure*}

Following \citet{Wang2022}, after barycentric and binary corrections with the ephemeris provided by \citet{CameroArranz2012}, we estimated the pulse period of each \textit{Insight}-HXMT observation using the epoch-folding method.
The pulse profiles were obtained by folding the background-subtracted light curves at given energy ranges with a phase bin of 32.
The maximum in each of the 24 pulse profiles was normalized to unity. The pulse profiles obtained for different observations were aligned according to an averaged template of 30-120\,keV (which is relatively simple and stable) with the {\sc FFTFIT} routine \citep{Taylor1992}
\footnote{We tried other alignment methods, such as using a sharp feature as the reference, and eventually obtained comparable results.}.
The evolution of pulse profiles has been shown in Figure~2 reported by \citet{Wang2022} and the detailed spin history is given in Table~2 reported by \cite{Hou2023arXiv230101423H}.
We note that the alignment may not be perfect due to variations in the pulse profile, and some phase offsets are expected in practice. We show all pulse profiles used in the following analysis in Figure \ref{profile1}.

\section{Decomposition analysis} \label{method}
Here we briefly summarize main assumptions and steps of the pulse profile decomposition analysis method proposed and comprehensively described by \citet{Kraus1995ApJ...450..763K}.
The basic assumption of the method is that observed asymmetric pulse profiles of X-ray pulsars can be represented as a combination of two symmetric (in phase) components associated with emission from regions around two magnetic poles. 
Each single-pole pulse profile is then a function of $\theta$, i.e., the angle between the magnetic axis and line of sight. 
It is symmetric with respect to two points $\Phi$ and $\Phi+\pi$, corresponding to the instant at which the pole is either closest or furthest from the direction of observation.
If the magnetic field is an ideal dipole field, both poles will have the same symmetry points, and thus the total pulse profile can only be symmetric as well. In general this is not, however, consistent with observations. So \citet{Kraus1995ApJ...450..763K} proposed a distorted magnetic dipole field, i.e., two magnetic poles are not located opposite to each other but rather offset by some angles.

\begin{figure*}[!htb]
\centering
\includegraphics[width=3.1in]{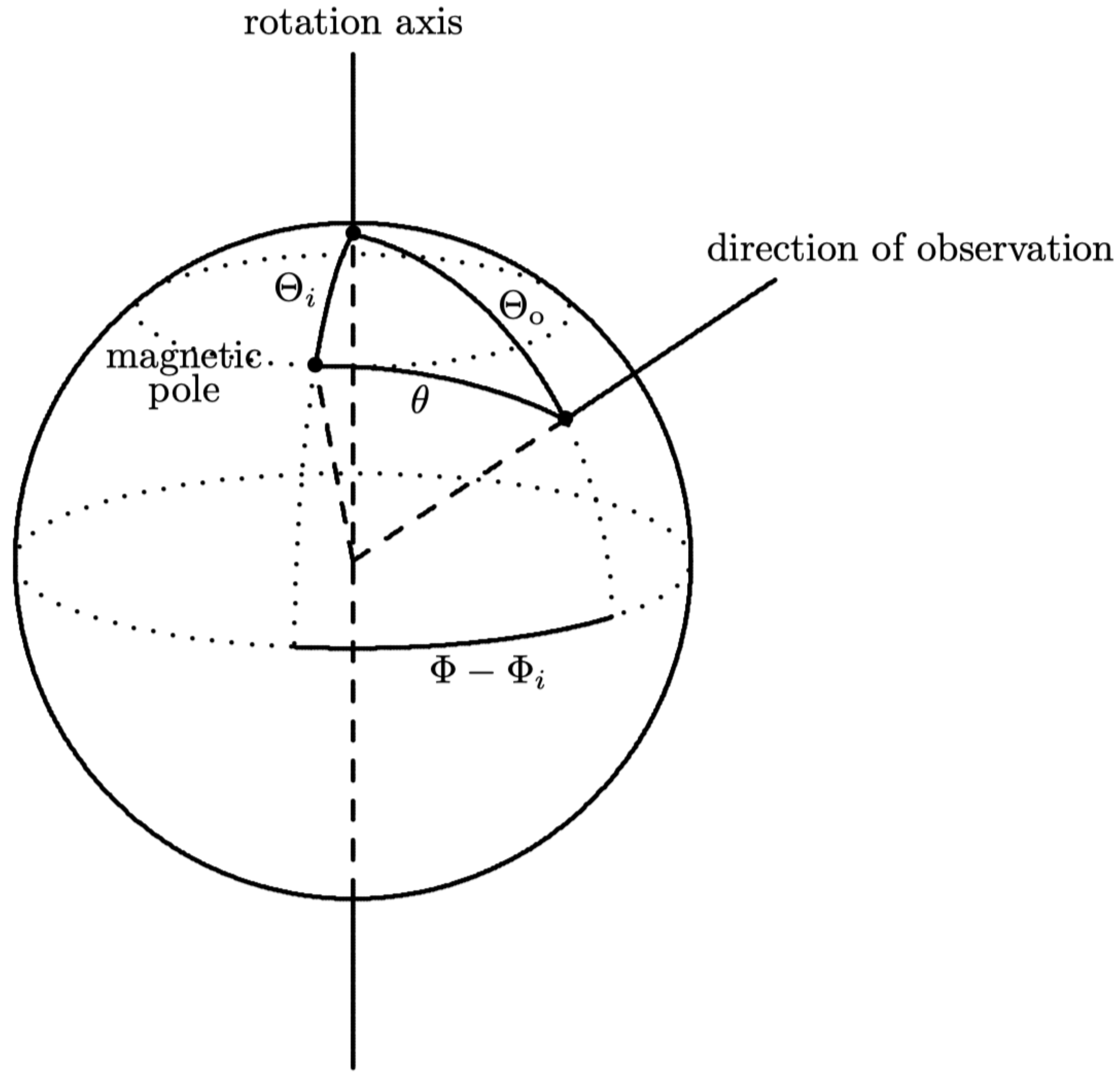}
\includegraphics[width=2.8in]{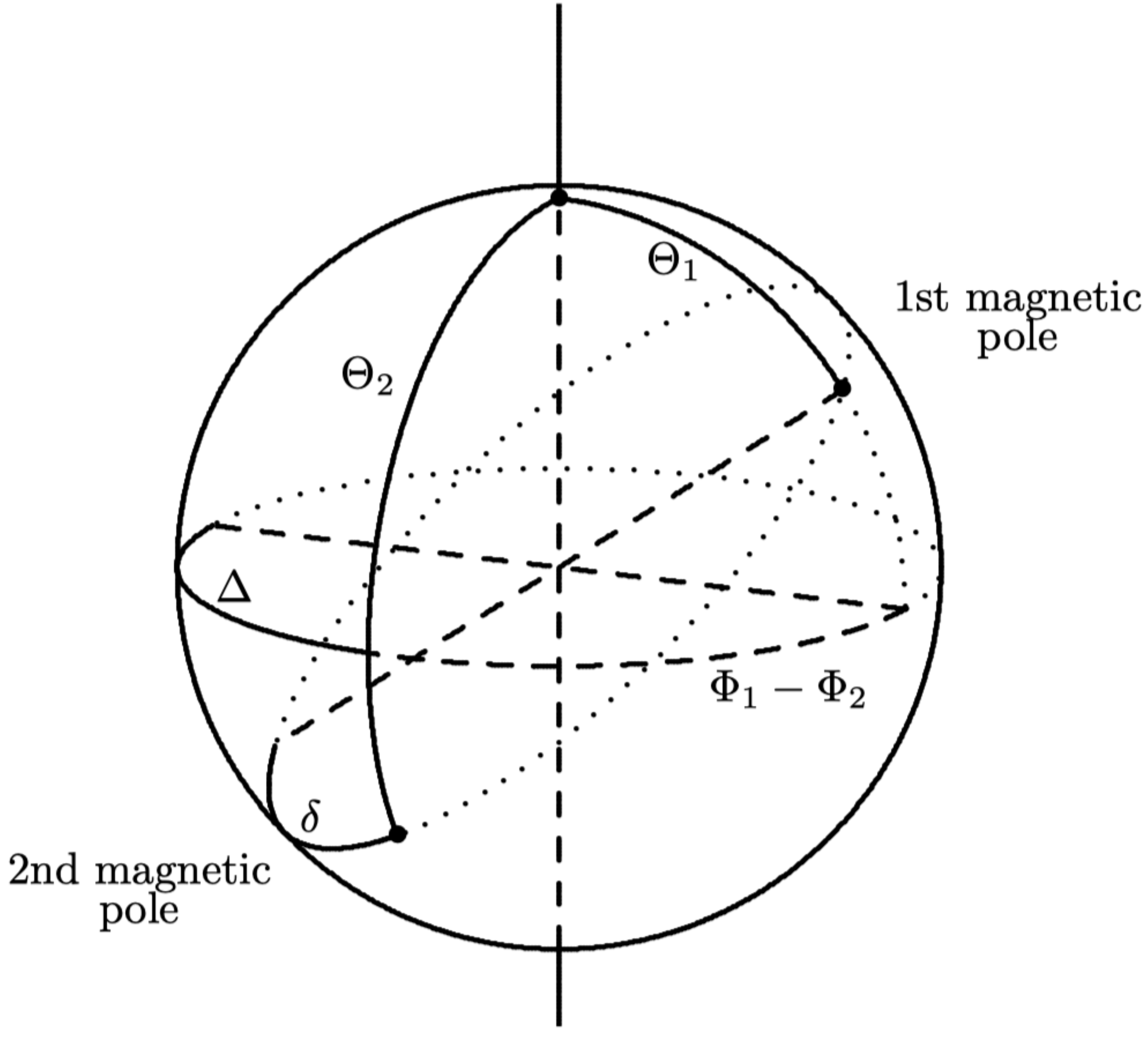}
\caption{Intrinsic geometry of the neutron star. Figures are adopted from \cite{Kraus1995ApJ...450..763K}.}
\label{jihe}
\end{figure*}

The corresponding basic geometry of the pulsar is shown in Figure~\ref{jihe}. $\Theta_{0}$ is the polar angle of the direction of observation. The magnetic poles are located at polar angles $\Theta_{1}$, $\Theta_{2}$. $\theta$ is the angle between a magnetic pole and the direction of observation, which is a function of the rotation phase $\Phi$. For each pole, the relation between $\theta$,$\Theta_{i}$,and $\Phi_{i}$ can be determined using the spherical triangles:
\begin{equation}
\cos\theta=\cos\Theta_{0}\cos\Theta_{i}+\sin\Theta_{0}\sin\Theta_{i}\cos(\Phi-\Phi_{i})
\label{theta}
\end{equation}  
The angular distance $\delta$ between one magnetic pole and the point that is antipodal to another magnetic pole represents the deviation from an ideal dipole magnetic field.
The corresponding difference of the azimuthal angle is $\Delta:=\pi-(\Phi_1-\Phi_2)$. 
The angular distance ($\delta$) between one magnetic pole and the point that is antipodal to the another magnetic pole can be written as:
\begin{equation}
\cos\delta=-\cos\Theta_{2}\cos\Theta_{1}+\sin\Theta_{2}\sin\Theta_{1}\cos\Delta.
\label{delta}
\end{equation}

\begin{figure*}[!htp]
\centering
  \includegraphics[width=0.8\textwidth]{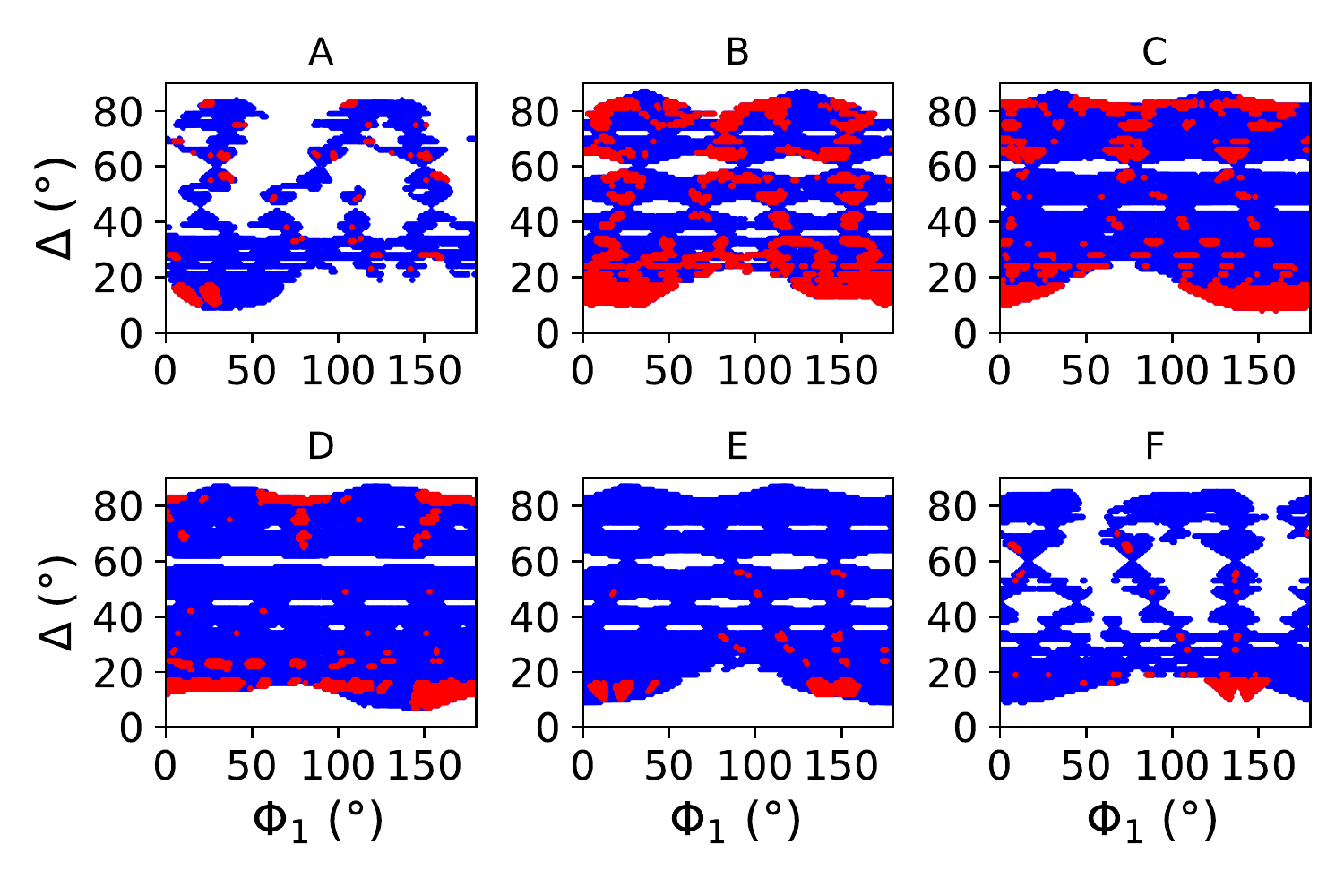}
\centering
\caption{Acceptable decompositions of $\Phi_1$ and $\Delta$ after applying the non-negative criteria (blue points) and non-negative plus no-ripples criteria (red points).}
\label{nonnegative}
\end{figure*}

\subsection{Decompositions}\label{Decomposition of pulse profiles}
As discussed by \citet{Kraus1995ApJ...450..763K}, it is convenient to search for possible decompositions in Fourier rather than real space. 
The observed pulse profile $F(\Phi)$ can be expressed as a Fourier series \citep{Kraus1995ApJ...450..763K}:
\begin{equation}
F(\Phi)=\frac{1}{2}u_{0}+\displaystyle\sum_{k=1}^{n/2-1} [u_{k}\cos(k\Phi)
+v_{k}\sin(k\Phi)]+u_{n/2}\cos(\frac{n}{2}\Phi),
\label{F}
\end{equation}
where $n$ is the number of bins of pulse profiles, $\Phi$ is the phase, and $u_{k}$, $v_{k}$ are coefficients that can be calculated by
\begin{eqnarray}
u_{k}=\frac{1}{\pi}\int_{-\pi}^{+\pi}F(\Phi)\cos(k\Phi)d\Phi,\\
v_{k}=\frac{1}{\pi}\int_{-\pi}^{+\pi}F(\Phi)\sin(k\Phi)d\Phi.
\label{uv}
\end{eqnarray}
As suggested by \citet{Kraus1995ApJ...450..763K}, we only considered the first 10 terms ($k \leq 10$) and omitted higher terms which are not really constrained by observations.
$F(\Phi)$ can be written as a sum of two single-pole pulse profiles $f_{1}(\Phi)$ and $f_{2}(\Phi)$, which are assumed to be symmetric with respect to points $\Phi_{1}$ and $\Phi_{2}$, respectively.
Therefore, their Fourier expansions can be written as
\begin{equation}
\vspace{-0.2cm}
f_{1}(\Phi)=\frac{1}{2}c_{0}+\displaystyle\sum_{k=1}^{n/2} c_{k}\cos[k(\Phi-\Phi_{1})],
\label{f1}
\end{equation}
and
\begin{equation}
f_{2}(\Phi)=\frac{1}{2}d_{0}+\displaystyle\sum_{k=1}^{n/2} d_{k}\cos\{k[\Phi-(\Phi_{2}+\pi)]\},
\label{f2}
\end{equation}
It can be shown that for arbitrary choice of symmetry points $\Phi_{1}$ and $\Phi_{2}$, coefficients in $f_{1}(\Phi)$ and $f_{2}(\Phi)$ , i.e., $c_{k}$ and $d_{k}$ ($k\neq0$), can be uniquely determined by solving $F(\Phi)=f_{1}(\Phi)+f_{2}(\Phi)$. On the other hand, the unmodulated flux ($u_{0}=c_{0}+d_{0}$) which represent the 0th frequency coefficient in fourier decomposition cannot be determined in this way. We first defined, therefore, the minimum values $c_{\rm 0, min}$ and $d_{\rm 0, min}$ of each single-pole pulse profile by shifting the minima of $f_{1}(\Phi)$ and $f_{2}(\Phi)$ to zero.
The distribution of the residual flux $u_{0}-c_{\rm 0, min} - d_{\rm 0, min}$ was estimated when combining two beam patterns (see below). We note that although the decomposition exists for every choice of $\Phi_{1}$ and $\Phi_{2}$, not all decomposition make sense \citep{Kraus1995ApJ...450..763K}. We selected, therefore, only the solutions that satisfy the following criteria:
\begin{enumerate}[(1)]
\item Non-negative: all values in $f_{1}(\Phi)$ and $f_{2}(\Phi)$ are non-negative because they represent the flux.
\item No-ripples: $f_{1}(\Phi)$ and $f_{2}(\Phi)$ are not expected to have small-scale features that cancel out in the sum. Also, the single-pole pulse profiles $f_{1}(\Phi)$ and $f_{2}(\Phi)$ should not be much more complicated than the observed total pulse profile.
\item Same geometry: the symmetry points $\Phi_{1}$ and $\Phi_{2}$ should be acceptable for different energy bands of all observations.
\end{enumerate}

\begin{figure*}
  \centering
  \includegraphics[width=0.8\textwidth]{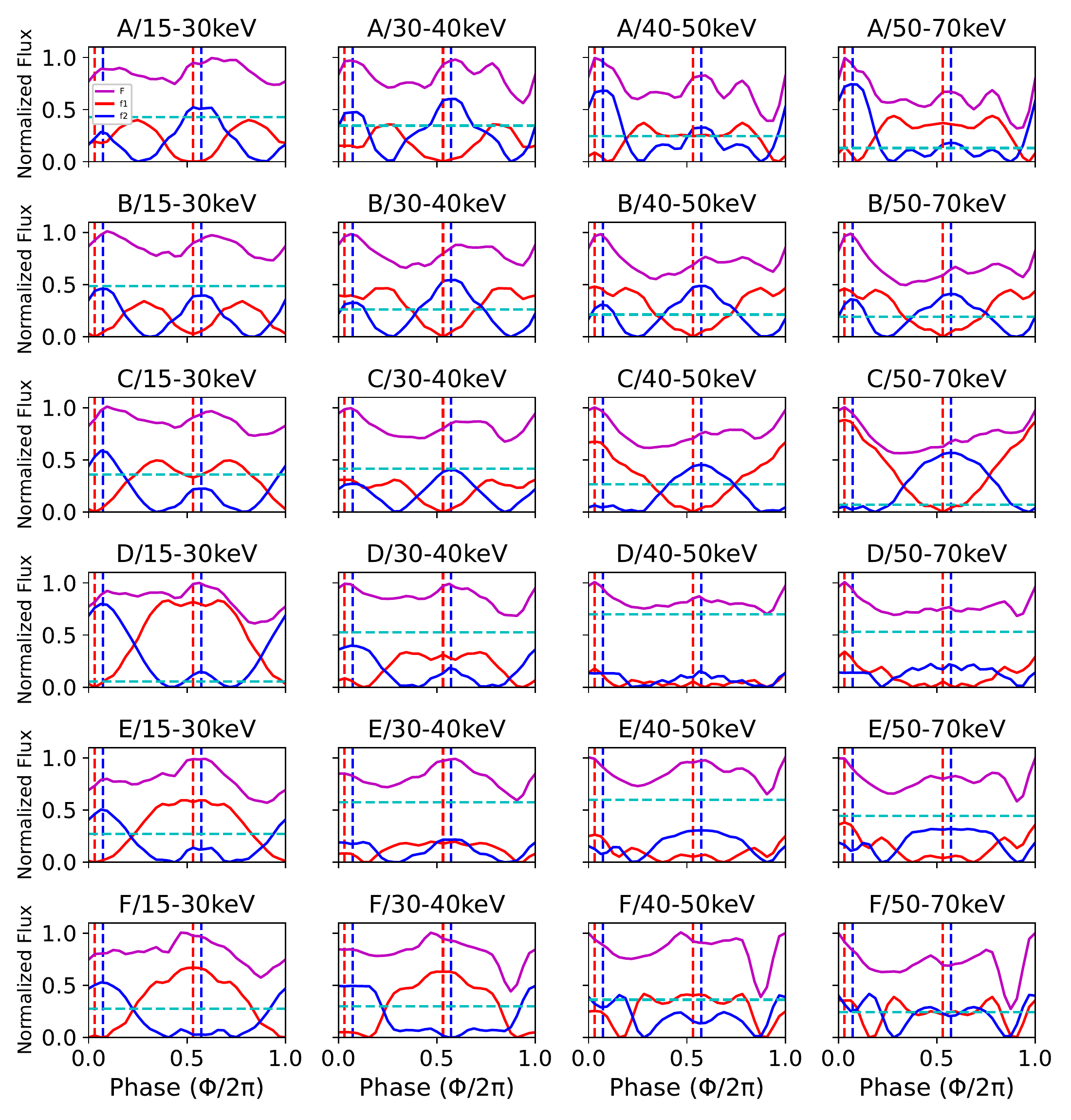}\vspace{-0.3cm}
\caption{ 
Decompositions of original pulse profiles $F(\Phi)$ (purple lines) into single-pole contributions $f_{1}(\Phi)$ and $f_{2}(\Phi)$ (red and blue lines) for all observations and energy bands. 
The unpulsed flux $u_{0}-c_{0,min}-d_{0,min}$ is shown by the cyan dotted horizontal lines.
Symmetry points of each single-pole pulse profile (i.e., $\Phi_{1}$, $\Phi_{1}+\pi$, $\Phi_{2}$ and $\Phi_{2}+\pi$) are indicated by dotted vertical lines.}
\vspace{0.1cm}
\label{profile2}
\end{figure*}

\subsection{Pulsar geometry}
As the pulsar rotates, the angle $\theta$ between one magnetic pole and the line of sight varies with the phase $\Phi$ in the range $\theta \in [\theta_{\rm min}, \theta_{\rm max}]$,
and the angle's range for another magnetic pole is $\theta \in[\theta'_{\rm min}, \theta'_{\rm max}]$, which may have an overlapping range with $[\theta_{\rm min}, \theta_{\rm max}]$.
$\theta_{\rm min}$, $\theta_{\rm max}$, $\theta'_{\rm min}$ and $\theta'_{\rm max}$ are related to the geometry of the pulsar, i.e., the polar angle ($\Theta_1$ and $\Theta_2$) of the magnetic pole and the viewing angle $\Theta_0$.
Following \citet{Kraus1995ApJ...450..763K}, we assume that intrinsic beam patterns from both poles should be the same even if different parts of each may be observed.
In this case, considering that the beam pattern from the two poles is only a function of $\theta$, it is possible to identify an overlapping region of the beam pattern from both magnetic poles and thus recover a larger fraction of the full beam pattern. 
Although this assumption is probably oversimplified\footnote{
Beam patterns at different magnetic poles might be not strictly identical because of possible asymmetric accretion flows, as suggested in 4U 1626–67 \citep{Iwakiri2019ApJ...878..121I}.}
, it was partially verified by overlapping regions found in Cen X-3 and Her X-1 \citep{Kraus1996ApJ...467..794K,Blum2000ApJ...529..968B}, which are consistent with recent polarization observations \citep{Doroshenko2022,Tsygankov2022}.
In practice, this means that at an instant $\Phi$ one pole is observed at an angle $\theta$, and the same angle is observed for the second pole at another instant $\tilde \Phi$ \citep{Kraus1995ApJ...450..763K}.
In this case, the relation between $\Phi$ and $\tilde \Phi$ is
\begin{equation}
\cos(\Phi-\Phi_{1})=\frac{\cot\Theta_{0}(\cos\Theta_{2}-\cos\Theta_{1})}{\sin\Theta_{1}}+\frac{\sin\Theta_{2}}{\sin\Theta_{1}}\cos(\tilde{\Phi
}-\Phi_{2}).
\label{Phi}
\end{equation}
For convenience, we write as
\begin{equation}
\cos(\Phi-\Phi_{1})=a+b\cos(\tilde{\Phi}-\Phi_{2}),~b>0,
\label{ab}
\end{equation}
where $a$ and $b$ can be estimated by minimizing the deviation of single-pole pulse profiles in the overlapping region of the beam pattern (for details, see Equ. 35 in \citet[][]{Kraus1995ApJ...450..763K}).
In this step, the distribution of the remaining flux $u_{0}-c_{\rm 0, min} - d_{\rm 0, min}$ is also calculated.
If the viewing angle $\Theta_0$ is known independently, the geometry of the pulsar can finally be determined by
\begin{equation}
\tan\Theta_{1}=\frac{-2a\,\tan\Theta_{0}}{(a\,\tan\Theta_{0})^2+b^2-1},
\label{Theta1}
\end{equation}
and
\begin{equation}
\tan\Theta_{2}=\frac{b\,\tan\Theta_{1}}{a\,\tan\Theta_{0}\,\tan\Theta_{1}+1}.
\label{Theta2}
\end{equation}

\section{Results}\label{results} 
To search for acceptable decompositions, we need to consider all possibilities in the ($\Phi_{1}, \Phi_{2}$) parameter space where $0 \leq \Phi_{1}, \Phi_{2} \leq \pi$.
For convenience, we replaced $\Phi_{2}$ with an auxiliary variable $\Delta:=\pi -(\Phi_{1}-\Phi_{2})$, which represents the azimuthal displacement between the two magnetic poles (see Figure~\ref{jihe}).
In this case, the searched parameter space becomes $0\le\Phi_{1}\le\pi$ and $0\le\Delta\le\pi/2$. 
We defined, therefore, grids with steps $1^{\circ}\times1^{\circ}$ covering this range, and tested  the above criteria one by one for each $\Phi_{1}$-$\Delta$ selection.

We first applied the non-negative criterion (blue points shown in Figure~\ref{nonnegative}), which means that the remaining flux $u_{0}-c_{\rm 0, min} - d_{\rm 0, min}$ should be positive when offsetting the minimum points of both single-pole profiles ($f_1$ and $f_2$) to zero (i.e. to ensure that observed flux from each of the poles is positive at all phases for pulse profiles in all energy bands and in all observations).
Figure~\ref{nonnegative} shows the decompositions that are acceptable for all energy bands of each observation (A-F).
For the no-ripples criterion, we estimated the complexity of pulse profiles by counting the number of peaks, calculated using the Python module Scipy\footnote{\url{https://docs.scipy.org/doc//scipy/reference/generated/scipy.signal.find_peaks.html}}.
Following \cite{Kraus1995ApJ...450..763K}, we excluded decompositions if their single-pole pulse profiles have much more peaks than the total pulse profile.  Figure~\ref{nonnegative} demonstrates acceptable solutions after applying both the non-negative and no-ripples criteria (red points).
We then searched for the common solution of A-F observations in order to satisfy the ``same geometry" criterion.
However, no such solution could be found.
We speculate that this might be due to the imperfect alignment between different observations, because of the variable shape of pulse profiles.  To account for such possibility, we assumed that there might be additional systematic error of 5 degrees related to imperfect alignment of pulse profiles from different observations as suggested by \cite{Caballero2011A&A...526A.131C} , and searched for possible decompositions again. Finally, some solutions were found, clustered around $\Phi_1=11^{\circ}/191^{\circ}$ and $\Phi_2=26^{\circ}/206^{\circ}$.
The corresponding $\Delta$ is $\sim$ $15^{\circ}$ if the dipole magnetic field is not dramatically distorted. We note that the two solutions are responsible for the two symmetry points (i.e., $\Phi_{\rm i}$ and $\Phi_{\rm i}+\pi$) for each single-pole pulse profiles. We can not decide which point corresponds to the instant at which the pole is closest to (or farthest from) the line of sight, so both possibilities were considered in the following, and we called them ``{\it plus} (+)" and ``{\it minus} (-)" solutions, respectively.
Selecting the more realistic solution shall then be done based on extra arguments such as comparison with theoretical pulse profile models (see below). The finally obtained single-pole pulse profiles are presented in Figure~\ref{profile2} together with the un-pulsed flux.

We finally calculated the beam pattern as seen by a distant observer according to the single-pole pulse profiles.
In practice, we first searched for the overlapping region of beam patterns based on the deviations between single-pole profiles for all observations and energy bands (for details, see Eq. 35 in \citet{Kraus1995ApJ...450..763K}).
However, unlike the cases in Her X-1 and Cen X-3 \citep{Kraus1996ApJ...467..794K, Blum2000ApJ...529..968B}, no overlapping region could be found, and therefore both $a$ and $b$ in Equ.~\ref{ab} could not be determined directly.
This is likely due to the fact that the geometry of the pulsar only allows the observer to see different sections of the total beam pattern. We assumed that $b$ is close to 1, which corresponds to the case where the magnetic field is not dramatically distorted \citep{Kraus1995ApJ...450..763K}.
On the other hand, following previous studies in EXO 2030+375 and 1A 0535+262 that show similar geometries \citep{Sasaki2010A&A...517A...8S, Caballero2011A&A...526A.131C}, $a$ was estimated to be around -2.2, based on the assumption that the sections of the two single-pole beam patterns can almost be connected to each other with a small gap.
In Figure~\ref{beampattern}, we show the reconstructed beam patterns for all observations and energy bands.
Here both the $\theta_{+}$ and $\theta_{-}$ represent the angle between the dipole magnetic axis and the line of sight to a distant observer, which correspond to the two possible solutions ({\it plus} and {\it minus}) as mentioned above.
The relation between the two solutions is $\theta_{+} =\pi- \theta_{-}$.

We calculated the polar angles of the pulsar using Eq.~\ref{Theta1} and \ref{Theta2}. For a given $\Theta_{0} = 37^{\circ}$ estimated from the orbital inclination \citep{Giovannelli2007A&A...475..651G}, the resulting $\Theta_{1}$ and $\Theta_{2}$ are $50^{\circ}$ and $130^{\circ}$, respectively. Therefore, the angular distance $\delta$ (between one magnetic pole and the point that is antipodal to the another magnetic pole) is $12^{\circ}$ according to Eq.~\ref{delta}. 
Considering the error propagation, the error of $\delta$ is about $3^{\circ}$ if typical errors of $\Delta$, $a$, $\Theta_{0}$ are assumed to be $5^{\circ}$, 0.1 and $2^{\circ}$, respectively.

\section{Discussion}\label{discussions}
Based on the extensive {\it Insight}-HXMT observations of the 2020 giant outburst of 1A~0535+262, we extracted energy-dependent pulse profiles at different luminosity states and decomposed them into single-pole contributions using the method proposed by \citet{Kraus1995ApJ...450..763K}.
We considered several physically motivated criteria to select reliable decompositions (see Sect.~\ref{Decomposition of pulse profiles}), and found that only solutions defined by symmetry points $\Phi_1=11^{\circ}/191^{\circ}$ and $\Phi_2=26^{\circ}/206^{\circ}$ are acceptable for all pulse profiles considered in our study.
The corresponding angle $\Delta$ defining offsett of the dipole from the center is found to be around $15^{\circ}$, which is slightly smaller than the previous estimates (i.e., $33^{\circ}\pm5^{\circ}$) inferred from different observations using the same method \citep{Caballero2011A&A...526A.131C}.
Since the $\Delta$ angle is a system parameter of the pulsar, it is not predicted to change significantly. Therefore, this deviation might reflect a systematic error of the decomposition method due to the imperfect underlying assumptions.
For example, recent studies reveal the presence of multi-pole magnetic fields which will also influence the observed pulse profiles \citep{Monkkonen2022, Kong2022}.
In addition, only in 1A~0525+262 there are independent studies based on different pulse profiles obtained from different outbursts. The validity of the decomposition method needs to be further tested, for instance through polarimetric observations of more X-ray pulsars.

\begin{figure*}
 \centering
   \includegraphics[width=0.8\textwidth]{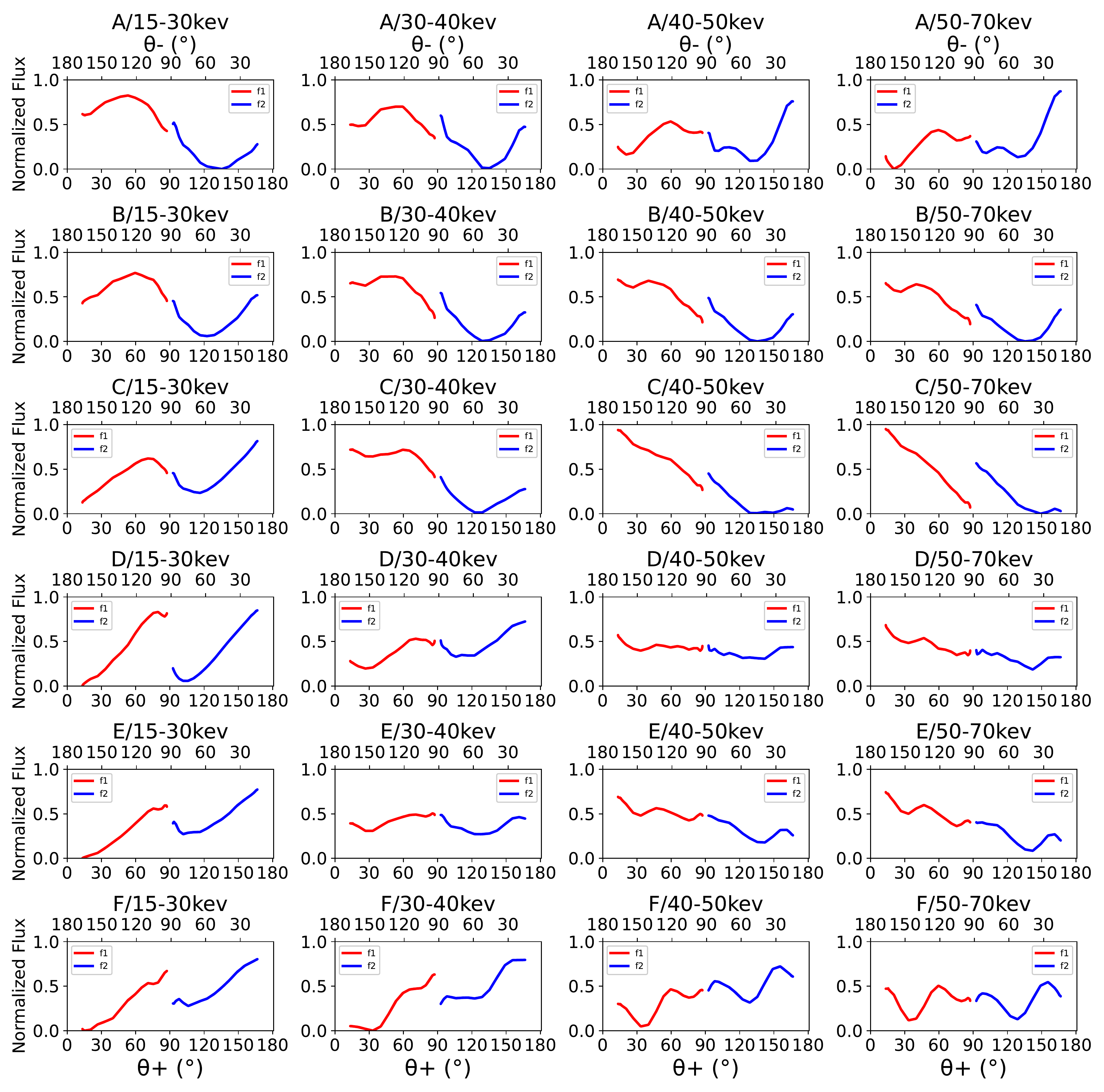}\vspace{-0.2cm}
 \caption{
 The reconstructed beam patterns of 1A 0535+262 for different observations and energy bands.
 The {\it plus} ($\theta+$) and the {\it minus} ($\theta-$) solutions correspond to the lower and upper x-axis, respectively.
 The red and blue lines represent different parts of beam patterns inferred from two single-pole pulse profiles.
 }
 \label{beampattern}
 \end{figure*}

Nevertheless, we divided the total pulse profiles into single-pole contributions (Figure~\ref{profile2}) and searched for overlapping regions of the beam pattern for the recovered geometry.
Eventually, no overlapping region was found, which is consistent with previous report by \citet{Caballero2011A&A...526A.131C}.
This suggests that the two single-pole profiles are responsible for different parts of the total beam pattern.
Similar results have been obtained in other sources, e.g., EXO 2030+375 \citep{Sasaki2010A&A...517A...8S}. 
We estimate the total beam pattern by assembling the two parts with a small gap (Figure~\ref{beampattern}).
In an ideal situation, it should be possible to connect the two parts by adjusting the distribution of the un-pulsed flux into the two single-pole profiles.
As shown in Figure~\ref{beampattern}, in most cases the two beam pattern parts can be connected although there are some exceptions, such as in 50-70\,keV in Observation C and in 15-30\,keV in Observation D.
The discontinuity of the beam pattern may be due to the fact that the gap we assumed is too small.
On the other hand, we cannot rule out the possibility that the beam pattern indeed changes suddenly because of the obscuration by the accretion column or the neutron star.
In 1A 0535+262, we found that the dipole magnetic field is not significantly distorted, which has a small offset $\delta \sim 12^{\circ}$ between one pole and the antipodal position of the other pole. This offset is slightly larger than that in Her X-1 \citep{Blum2000ApJ...529..968B}, and is comparable to that in Cen X-3 \citep{Kraus1996ApJ...467..794K}
\footnote{In literature, large offsets were reported in 4U 0115+63, V 0332+53 and EXO 2030+375 \citep{Sasaki2010A&A...517A...8S, Sasaki2012A&A...540A..35S}. However, these results were based on unknown and assumed viewing angles, leading to large uncertainties of the resulting $\delta$ (e.g., see Figure~5 in \citet{Sasaki2012A&A...540A..35S}).
}.
Recent studies indicate that the geometries of Her X-1 and Cen~X-3, inferred from the pulse profile decomposition, are consistent with polarization observations \citep{Doroshenko2022,Tsygankov2022}. Therefore, we strongly encourage polarization studies of 1A 0535+262 during the future giant outbursts that occurs every a few years with observatories such as \textit{Imaging X-ray Polarimeter Explorer} \citep{Weisskopf2022} and \textit{enhanced X–ray timing and polarimetry} \citep{Zhang2019}.

\begin{figure*}
  \centering
  \includegraphics[width=0.78\textwidth]{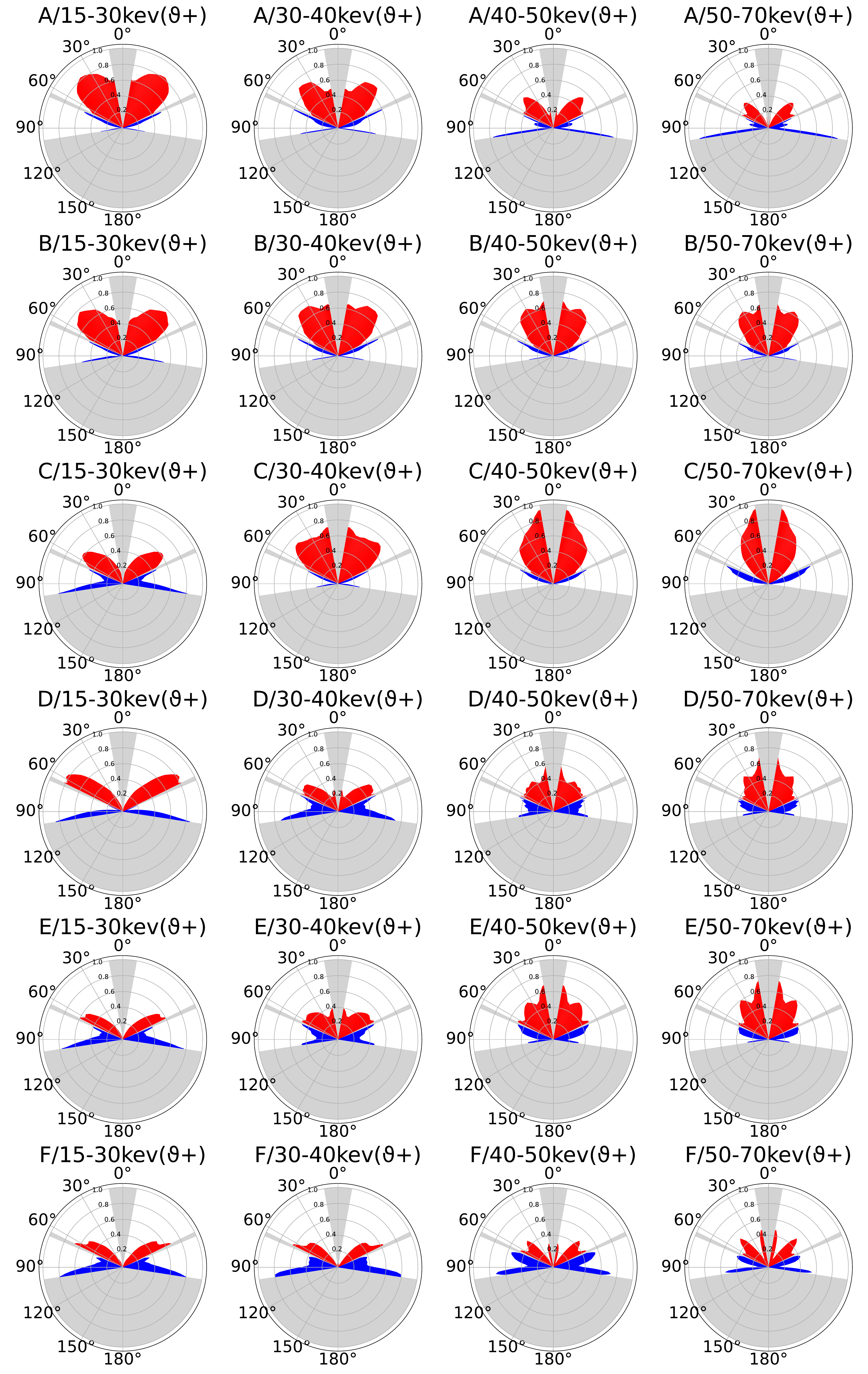}\vspace{-0.3cm}
\caption{Intrinsic beam patterns in polar representation for different observations and energy bands considering the {\it plus} solutions, where the angle $\vartheta$ presents the angle between the dipole magnetic axis and the normal to the stellar surface measured by a local observer in the comoving frame.
The radial direction indicates the intensity of the beam pattern. 
The red and blue shaded regions correspond to the first and second magnetic poles, and the gray region shows the angles where no information is available.}
\vspace{0.1cm}
\label{intrinsicbeampattern1}
\end{figure*}

\begin{figure*}
  \centering
  \includegraphics[width=0.78\textwidth]{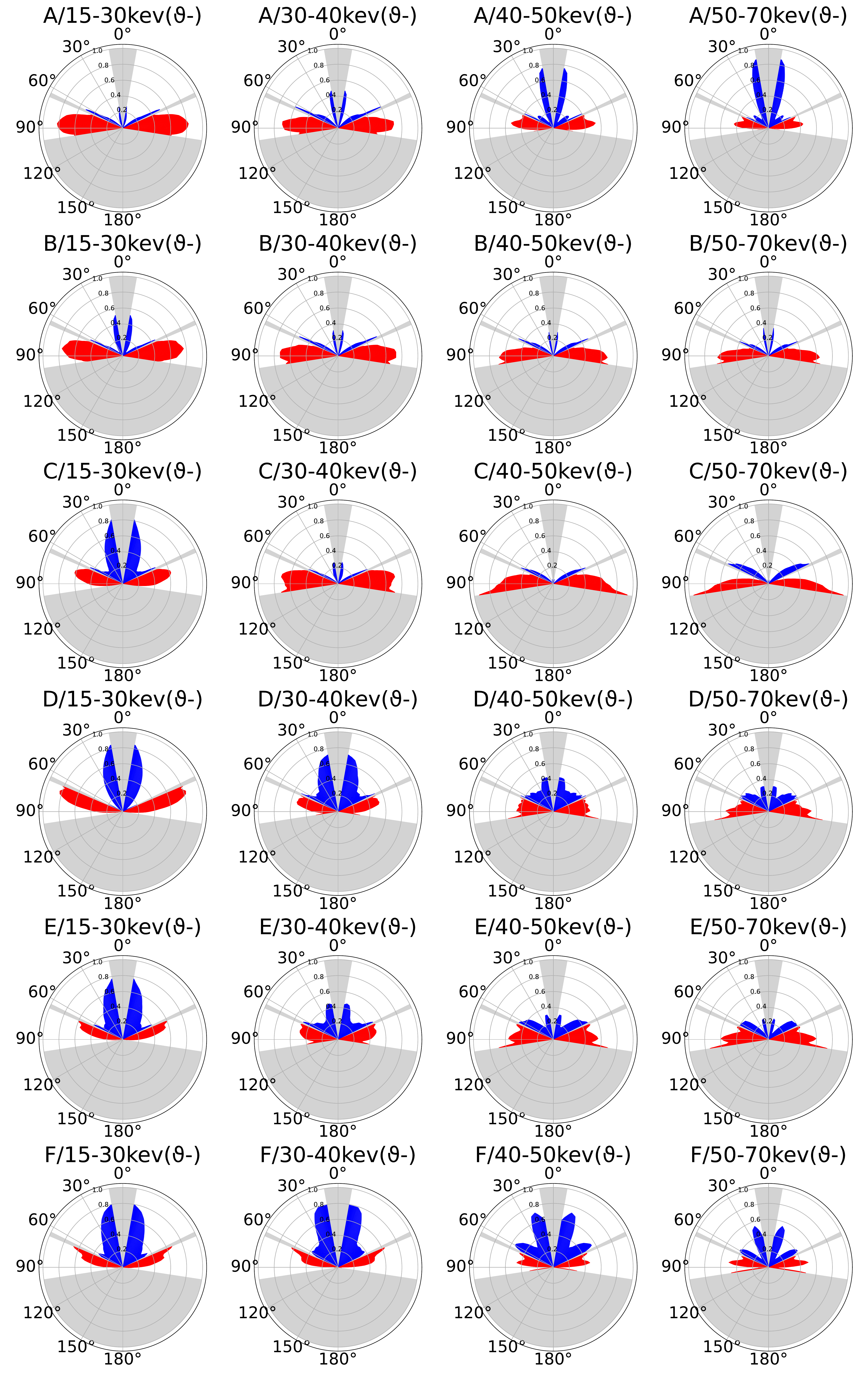}\vspace{-0.3cm}
\caption{Intrinsic beam patterns in polar representation for different observations and energy bands considering the {\it minus} solutions, where the angle $\vartheta$ presents the angle between the dipole magnetic axis and the normal to the stellar surface measured by a local observer in the comoving frame.
The radial direction indicates the intensity of the beam pattern.
The red and blue shaded regions correspond to the first and second magnetic poles, and the gray region shows the angles where no information is available.}
\vspace{0.1cm}
\label{intrinsicbeampattern2}
\end{figure*}

It is known that the relativistic light bending has a significant effect on observed pulse profiles.
The specific radiation region of pulsars, for instance, the height of the accretion column, is still poorly known.
If we assume that the radiation is mainly emitted around polar caps and on the surface of the neutron star, we can convert the apparent beam pattern to the intrinsic beam pattern (shown in Figure~\ref{intrinsicbeampattern1} and Figure~\ref{intrinsicbeampattern2}) using the approximate formula \citep{Beloborodov2002},
\begin{equation}
{\rm cos}\vartheta \approx {\rm cos\theta}(1-\frac{r_{\rm g}}{R}) + \frac{r_{\rm g}}{R},
\end{equation}
where $\vartheta$ is the angle between the radiation direction and the normal to the stellar surface measured by a local observer in the comoving frame.
$r_{\rm g}$ and $R$ are the Schwarzschild radius and the radius of the neutron star, which has $R=2.4\,r_{\rm g}$ considering canonical values, i.e., $R$=10\,km and the mass of the neutron star $M=1.4M_{\odot}$.

The basic picture of the accretion process onto highly magnetized neutron stars has been proposed by many authors \citep[e.g.,][]{Nagel1981, Meszaros1992}.
However, it is still difficult to reproduce the pulse profiles in theory.
This is due to the fact that many non-linear effects need to be taken into account, most notably strong energy and magnetic field dependent scattering cross-sections defining plasma opacities and thus radiative pressure and the dynamical structure of the accretion flow, the gravitational light bending \citep[for details, see][]{Falkner2018}, and more.
Generally, the radiation we observed is emitted from the accretion mound/column directly or the reprocessing via the surface of the neutron star and/or the upper accretion stream.
It is generally accepted that at low luminosities there is an accretion mound on the polar cap of the neutron star, and the emission is mainly transported (and scattered) through the infalling matter, forming a ``pencil" beam parallel to the magnetic field lines. 
On the other hand, at high luminosities, an accretion column appears. As a result, photons can only escape from the sides of the column and perpendicularly to the magnetic field, leading to a ``fan" beam  \citep{Basko1976MNRAS.175..395B, Becker2012}.
As shown in Figure~\ref{intrinsicbeampattern2}, the {\it minus} solution has indeed two main components parallel and perpendicular to the magnetic field (i.e., $\vartheta \sim 0^{\circ}/90^{\circ}$) respectively, which  mimics the combination of the canonical ``pencil" and ``fan" beam patterns, and therefore is more consistent with theoretical expectations (albeit rather simplistic).
In addition, the {\it minus} solution is similar to the beam patterns in Her X-1 and Cen X-3 \citep{Blum2000ApJ...529..968B, Kraus1996ApJ...467..794K}, which also suggests that it is probably the correct one for the pulsar.

\citet{Kong2021ApJ...917L..38K} studied the evolution of cyclotron resonant scattering features (CRSFs) in 1A 0535+262 during its 2020 giant outburst and found that the CRSF energy is positively (negatively) correlated with luminosity when the luminosity is smaller (larger) than a critical value $\rm 6.7\times10^{37} ergs/s$. 
This theoretically suggests the transition of accretion regimes between ``pencil" and ``fan" beam patterns. 
However, as shown in Figure~\ref{intrinsicbeampattern1} and Figure~\ref{intrinsicbeampattern2}, the beam pattern is more complex and energy-dependent.
The cyclotron line energy $E_{\rm cyc}$ is $\sim$ 45\,keV in 1A~0535+262 \citep{Kong2021ApJ...917L..38K}, resulting in dramatic changes of the cross section around this energy range and therefore significant variations of pulse profiles \citep{Wang2022}.
For the energy band of $E \gtrsim E_{\rm cyc}$ (i.e., 40-50\,keV and 50-70\,keV), the beam evolution is qualitatively consistent with theoretical expectations aforementioned, i.e., dominated by the ``pencil" beam when the source is relatively faint and dominated by the ``fan" beam around the outburst peak.
In Observations C, D and E when the source is bright, we find that there is a significant fraction of high energy X-rays emerging from the direction $\vartheta >90^{\circ}$. 
We consider that this might originate from the scattering in the upper accretion stream as suggested by \citet{Kraus2003, Sasaki2010A&A...517A...8S, Caballero2011A&A...526A.131C, Sasaki2012A&A...540A..35S}.
On the other hand, the beam pattern of $E < E_{\rm cyc}$ is more complex.
For example, in Observation A the 15-30\,keV pulse profile presents a ``fan" beam which is different from that of high energies.
To our knowledge, this is the first time the transition of beam patterns with energy is discovered.
This is consistent with the theoretical prediction by \citet{Brainerd1991}, who interprets it as a result of the scattering in the accretion column if the column is optically thin to Thomson scattering and optically thick to resonant Compton scattering.
In addition, another ``pencil" beam component also appears for pulse profiles at low energies (15-30\,keV and 30-40\,keV).
It is stronger in the fading phase of the outburst than that of the rising phase, even though the accretion rate is the same in both cases. This might be related to the hysteresis effects of spectral and temporal properties reported by other authors \citep[e.g.,][]{Doroshenko2017, Wang2020, Kong2021ApJ...917L..38K}.  The physical mechanism is still poorly known. Nevertheless, we speculate that this ``pencil" beam must be attributed to an accumulated effect, such as a gradual change of the shape of the accretion mound/column, which may influence the velocity of the in-falling matter near the accretion column's wall and therefore the illumination onto the surface of the neutron star.
As a result, the reflection \citep{Lyubarskii1988,Poutanen2013,Kylafis2021} might be stronger in the fading phase of the outburst, corresponding to the additional ``pencil" beam.

\section{acknowledgments}
This work is based on observations with {\it Insight}-HXMT, a project funded by the China National Space Administration (CNSA) and the Chinese Academy of Sciences (CAS). This work is supported by the National Natural Science Foundation of China under grants No. 12173103, U2038101, U1938103, 11733009. This work is also supported by International Partnership Program of Chinese Academy of Sciences (Grant No.113111KYSB20190020), the National SKA Program of China (Grant No. 2022SKA 0120101) and the National Key R$\&$D Program of China (No. 2020YFC2201200), the science research grants from the China Manned Space Project (No. CMSCSST-2021-B09, CMSCSST-2021-B12 and CMS-CSST-2021-A10), and opening fund of State Key Laboratory of Lunar and Planetary Sciences (Macau University of Science and Technology) (Macau FDCT Grant No. SKL-LPS(MUST)-2021-2023). C.Y. has been supported by the National Natural Science Foundation of China (Grant Nos. 11521303, 11733010, and 11873103).

\vspace{5mm}

\bibliography{sample631}{}
\bibliographystyle{aasjournal}

\end{document}